\documentclass[preprint,aps,showpacs,superscriptaddress,groupedaddress,nofootinbib]{revtex4-2}
\usepackage{graphicx}
\usepackage{dcolumn}   
\usepackage{bm}   
\usepackage{amssymb}
\usepackage{color}
\usepackage{amsmath}
\usepackage{mathrsfs}
\usepackage{latexsym}
\usepackage{natbib}    
\usepackage[toc,page]{appendix}
\usepackage{braket}
\newcommand{\del}{\partial}

\begin{document}

\title{Cosmological Consequences of Varying Couplings in Gravity Action}

\author{Sandipan Sengupta}
\email{sandipan@phy.iitkgp.ac.in}
\affiliation{Department of Physics, Indian Institute of Technology Kharagpur, Kharagpur-721302, INDIA}

\begin{abstract}
We develop a Lagrangian formulation for gravity with matter where the gravitational couplings are universally treated as being field-dependent. The solutions for FLRW geometries and the associated time evolution of the Newton and cosmological couplings are found. 
The distance-redshift relations are shown to prefer a slowly growing Newton's coupling alongwith a negative equation of state ($w\geq -\frac{1}{3}$) for the matter fluid at the present epoch of accelerated expansion, while ruling out the evolution of $G$ predicted by Dirac's large number hypothesis. We obtain an improved bound on $\dot{G}(t)$ as: $1.67\times 10^{-11} yr^{-1}<\frac{\dot{G}}{G}<3.34\times 10^{-11} yr^{-1}$ in the context of supernova cosmology, as well as a new constraint on $\dot{\Lambda}(t)$ as: $-0.67\times 10^{-11} yr^{-1}<\frac{\dot{\Lambda}}{\Lambda}<-1.34\times 10^{-11} yr^{-1}$. Based on this formulation, we also present a dynamical solution to the `cosmic coincidence' problem.

\end{abstract}

\maketitle
\newpage

\section{Introduction}

The question as to whether the couplings in the gravity action are constants or are varying in time is one of fundamental importance. Early attempts exploring the idea of a time-dependent Newton's coupling could be traced back to Dirac's large number hypothesis \cite{dirac,dirac1}, where the precise form of the variation was invoked towards explaining the origin of large dimensionless numbers ($\sim 10^{39}$) in nature. Later on, Brans and Dicke had proposed a gravitational action underlying a variable $G$ \cite{bd} with an aim to connect gravity to Mach's principle. The possibility of varying couplings have since been considered from various perspectives \cite{peebles,wetterich,shapiro,smolin2,smolin,*smolin1}. Phenomenological models of running couplings have also been richly investigated \cite{overduin,ujan,*uzan,Sola}.

However, the problem of formulating a universal Lagrangian framework where all possible gravitational couplings could vary (in time) requires a fresh outlook. Here, we set up a Lagrangian formulation for gravity where the possible time-evolution of both Newton's and cosmological coupling ($G$ and $\Lambda$, respectively) could be described through a single scalar field without introducing any modification (or, any potential) in the matter sector. Based on the resulting dynamics, we analyze the possible cosmological consequences in detail. 
The resulting theory is not the same as the one where the couplings are treated as constants from the outset, namely, before varying the action. The cosmological implications found here goes beyond the scope of standard FLRW cosmology and the $\Lambda CDM$ model, laying down rich future prospects.

Our analysis unravels the class of possible variations of couplings consistent with the field equations. These solutions should act as the guiding principles underlying potential phenomenological models in this context. Some of the consequences are rather radical and interesting. For instance, we show that for varying couplings as introduced within the proposed Lagrangian formulation, an accelerated expansion is possible for an equation of state $w\geq -\frac{1}{3}$, which includes the era of dust-matter domination.
There is also the possibility of a dynamical vanishing of $\Lambda(t)$, thus defining a critical epoch. Based on the distance-redshift diagram, we present a more stringent constraint on the variation of $G(t)$ than is currently available in supernova cosmology, and a new (to the best of our knowledge) bound for $\dot{\Lambda}(t)$. As another remarkable dynamical consequence, we find a natural solution to the well-known `cosmic coincidence' problem \cite{zlatev,manohar}.

In the next section, we introduce the Lagrangian density for gravity with matter with field-dependent couplings. The cosmological solutions and their general features are presented in the next two sections. The distance-redshift relation and its connection to current observations are then discussed. This is followed by a study of the implications for the coincidence problem. The concluding section contains a few relevant remarks.

\section{Lagrangian density with varying couplings}

Let us introduce the following first-order Lagrangian density for gravity in four dimensions:
\begin{eqnarray}\label{S}
{\cal L}[e,\omega,\xi,\psi]~=~\frac{\xi}{2} e e^{\mu}_{I} e^\nu_J {R}_{\mu\nu}^{~~IJ}(\omega)-e\lambda(\xi)~+~{\cal L}_m [\psi]
\end{eqnarray}
where $\xi$ is a scalar, $e_\mu^I$ are the tetrad, $\omega_\mu^{~IJ}$ are the spin-connection defining the field-strength as: $R_{\alpha\beta}^{~~IJ}(\omega)=\del_{[\alpha} \omega_{\beta]}^{~IJ}+\omega_{[\alpha}^{~IK}\omega_{\beta]K}^{~~~~J}$ and $\psi$ denote the matter field(s).  Note that both the Newton's and cosmological coupling are dependent on the single field $\xi$. The reason for not treating $\lambda$ as yet another independent field is simply that such a scenario necessarily forces the spacetime metric to be degenerate, a feature that is impractical and not desirable. Also, we do not include quadratic or higher curvature terms in the action, since our purpose is to keep as close as possible to Einstein gravity modulo the space of couplings.

The fact that ${\cal L}_m$ is independent of $\xi$ implies that the equation(s) of motion for the matter field(s) $\psi$ remain unaffected in comparison to Einstein theory. The gravitational field equations, on the other hand, are modified, reading:
\begin{eqnarray}\label{eom1}
&\delta e:~&\xi\big[R_{\alpha\beta}(\omega)~-~\frac{1}{2}g_{\alpha\beta}R(\omega)\big]~+~ g_{\alpha\beta}\lambda(\xi)~=~ T_{\alpha\beta},\nonumber\\
&\delta \omega:~& D_{\alpha}(\omega)\big[\xi e e^{[\alpha}_I e^{\beta]}_J\big]~=~0,\nonumber\\
&\delta \xi:~& R(\omega)~-~2\lambda'(\xi)~=~0,
\end{eqnarray}
where the Ricci tensor $R_{\alpha\beta}(\omega)$ and the Ricci scalar $R(\omega)$ are built from the connection with torsion, $T_{\alpha\beta}$ is the matter stress-energy tensor and $D_\mu(\omega)$ is the gauge-covariant derivative with respect to the connection. 
The solution to the torsion equation may be written in terms of the torsionless connection $\bar{\omega}_\mu^{~IJ}(e)$, given completely by the tetrad, as: $\omega_\mu^{~IJ}-\bar{\omega}_\mu^{~IJ}(e)\equiv K_\mu^{~IJ}=\frac{1}{2\xi}e^{\sigma [J}e_\mu^{I]}\del_\sigma \xi$. 

Note that $\frac{\lambda(\xi)}{\xi}\equiv\Lambda(\xi)$ emerges as the effective cosmological coupling. 
Importantly, the first and last equations in (\ref{eom1}) imply that this coupling is related to the trace $T$ of the energy-momentum tensor:
\begin{eqnarray}
2\big[2\lambda(\xi)~-~\xi \lambda'(\xi)\big]~=~T
\end{eqnarray}
This interesting feature may be contrasted with Einstein gravity. 
As the above equation reflects, this formulation is expected to have a bearing on the cosmological coincidence problem, an issue that would be analyzed after presenting some general results.

\section{FLRW cosmology: Field equations}

To analyse the possible cosmological consequences of the action (\ref{S}), we adopt the simplest possible setting, namely the spatially flat FLRW geometries ($ c=1$):
\begin{eqnarray*}
ds^2~=~-dt^2+a^2(t)[dr^2+r^2(d\theta^2+\sin^2 \theta d\phi^2)],
\end{eqnarray*}
The spatial homogeneity and isotropy implies that the field $\xi$ could depend only on time: $\xi=\xi(t)$. The matter sector as a whole, which could in principle consist of more than one matter fields, is assumed to admit a perfect fluid representation: $T_{\mu\nu}=(\rho+P)u_\mu u_\nu+Pg_{\mu\nu}$  where $u^\mu$ is the four-velocity of the fluid. 

With the above assumptions, the set of gravitational equations of motion read:
\begin{eqnarray}
&&\frac{3}{a^2}\Big[\dot{a}+\frac{a\dot{\xi}}{2\xi}\Big]^2~-~\frac{\lambda(\xi)}{\xi}~=~\frac{\rho}{\xi}\label{c-eom}\\
&& \frac{2}{a}\del_t\Big[\dot{a}+\frac{a\dot{\xi}}{2\xi}\Big]~+~\frac{1}{a^2}\Big[\dot{a}+\frac{a\dot{\xi}}{2\xi}\Big]^2~-~\frac{\lambda(\xi)}{\xi}~=~-\frac{P}{\xi}\label{c-eom1}\\
&& \frac{\lambda'(\xi)}{2}~-~\frac{\lambda(\xi)}{\xi}~=~\frac{\rho-3P}{4\xi}\label{c-eom2}
\end{eqnarray}
The set of three equations above specifies the complete system (for a given matter field), which is underdetermined. In order to solve for $a(t),\xi(t),\lambda(t),\rho(t),P(t)$, we need two additional conditions. Here we would solve the system by assuming a power law ansatz and a constant equation of state $w\equiv\frac{P}{\rho}=const.$ Since the standard FLRW solutions in Einstein theory with a constant $w$ emerge as power law, this makes a direct comparison of the resulting dynamics in the two cases possible. 

Taking a time derivative of eq.(\ref{c-eom}) and then using eqs.(\ref{c-eom1}) and (\ref{c-eom2}), we obtain:
\begin{eqnarray}\label{rhodot}
\dot{\rho}~=~-\frac{3\dot{a}}{a}(\rho+P)
\end{eqnarray}
Remarkably, the derivatives of $\xi$ and $\lambda(\xi)$ conspire to produce exactly the same local conservation law as in standard FLRW with constant couplings. Thus, although the field equations and hence the solutions $a(t)$ in general are different for varying couplings, the solution for the matter density $\rho(a)$ for a given $w$ turns out to be exactly the same as in standard cosmology: $\rho a^{3(1+w)}=const.$

Further, eq.(\ref{c-eom2}) shows that radiation defines a critical epoch, where the right hand side vanishes exactly. Hence, the cases $w\neq \frac{1}{3}$ and $w=\frac{1}{3}$ would be analyzed separately in the next two sections.

\section{Solutions for $w\neq \frac{1}{3}$}
Let us assume the following power law ansatz:
\begin{eqnarray}\label{p-law}
\xi=\xi_0\Big(\frac{t}{t_0}\Big)^{\alpha},~\lambda=\lambda_0\Big(\frac{t}{t_0}\Big)^{\beta}
\end{eqnarray}
The constants are related to the current values (at $t=t_0$) of $G$ and $\Lambda$ as: $\xi_0=\frac{1}{8\pi G_0}>0,~\lambda_0=\frac{\Lambda_0}{8\pi G_0} $.
Since we have introduced one more condition than required to solve the system of three equations, only two among the three parameters $w,\alpha$ and $\beta$ are independent. 

Eq.(\ref{c-eom2}) then leads to:
\begin{eqnarray}
\frac{\rho}{\xi}~=~\frac{2(\beta-2\alpha)}{(1-3w)\alpha}\frac{\lambda_0}{\xi_0}\Big(\frac{t}{t_0}\Big)^{\beta-\alpha}
\end{eqnarray}
Using this along with eqs.(\ref{c-eom}) and (\ref{c-eom1}) and assuming $\lambda\neq 0$, we obtain:
\begin{eqnarray*}
\frac{\dot{a}}{a}~=~-\frac{\beta}{3(1+w)t}
\end{eqnarray*}
whose solution is:
\begin{eqnarray}
a(t)~=~a_0 \Big(\frac{t}{t_0}\Big)^{-\frac{\beta}{3(1+w)}}
\end{eqnarray}
Finally, inserting this back into eq.(\ref{c-eom}), we obtain the following constraint, as anticipated at the beginning of this section:
\begin{eqnarray}
\alpha-\beta~=~2
\end{eqnarray}
where $\beta<0$ in an expanding universe.

\subsection{Generic features}

Let us emphasize a few essential details regarding the solutions just obtained:\\
a. For $w<\frac{1}{3}$, positivity of density requires: $\frac{(\alpha +2)\lambda}{\alpha}<0$. Hence, for a positive $\lambda$ the  Newton's coupling $\xi^{-1}\equiv G$ must grow in time: $-2<\alpha<0$. For $\lambda<0$, it could be either decaying ($0<\alpha<2$) or growing ($\alpha<-2$), corresponding to a slower and faster variation, respectively. 
The ultra-relativistic equations of state $w>\frac{1}{3}$, on the other hand,  exhibit exactly the opposite behaviour. \\
b. Under the assumption of a slow variation of $G(t)$, the only integer solutions for a decaying and growing $G(t)$ is $\alpha=1$ and $\alpha=-1$, respectively. For rational solutions, there exists an infinite number of them. However, the correct exponent could only be inferred from observations.\\
c. For both classes of solutions, $|\Lambda(t)|\sim t^{-2}$ exhibits a decaying behaviour.\\
d. An accelerated expansion of the universe requires $\frac{2-\alpha}{3(1+w)}>1$.
 Thus, for both $w>\frac{1}{3}$ and $w<\frac{1}{3}$, acceleration is possible. These two limits are relevant for the early and late universes, respectively.\\
e. The trivial case $\alpha=0$ ($G\equiv$ const.) reproduces the standard FLRW solutions.

\subsection{Examples}

{\bf A. Stiff fluid:}

For $P=\rho$, $a\sim t^{\frac{2-\alpha}{6}},~\rho\sim a^{-6}$. The two classes of solutions are defined by $\lambda>0$, associated with a slowly decaying ($2>\alpha>0$) or a rapidly growing  ($\alpha<-2$) $G(t)$, and $\lambda<0$ with a slowly growing ($-2<\alpha<0$) $G(t)$. This stiff
phase could dominate only in the early universe.
\vspace{.6cm}

{\bf B. Dust:}

For $P=0$, $a\sim t^{\frac{2-\alpha}{3}},~\rho\sim a^{-3}$. Note that for positive $\lambda$, solutions now correspond to only growing $G$ ($-2<\alpha<0$), whereas negative $\lambda$ is associated with either $2>\alpha>0$ (decaying) or $\alpha<-2$ (growing). Remarkably, the universe could accelerate even in the matter dominated era for positive $\lambda$ provided $\alpha<-1$. 
Assuming a positive $\lambda$, the age of a matter-dominated universe is given by: $\frac{2}{3H_0}\leq t_0<\frac{4}{3H_0}$. 
\vspace{.6cm}

{\bf C. Curvature fluid:}

As a prototype of a negative equation of state, let us consider $w=-\frac{1}{3}$, which also characterizes the spatial curvature term in standard FLRW case. The solutions read: $a\sim t^{\frac{2-\alpha}{2}},~\rho\sim a^{-2}$. While the qualitative features of the positive and negative $\lambda$ solutions are similar to the dust solutions, we note that $\ddot{a}>0$ is possible for $\lambda>0$ for any arbitrarily slow variation of $G(t)$ as long as it grows in time ($\alpha<0$). The associated limits on the age of universe dominated by this equation of state is: 
$\frac{1}{H_0}\leq t_0<\frac{2}{H_0}$ for $\lambda>0$. 

\section{Solutions for radiation matter}

For $w=\frac{1}{3}$, eq.(\ref{c-eom2}) is solved as:
\begin{eqnarray}
\lambda(\xi)~=~C\xi^2, ~C~\equiv~\mathrm{const.}
\end{eqnarray}
which now implies: $\beta=2\alpha$. The remaining eqs.(\ref{c-eom}) and (\ref{c-eom1}) then read:
\begin{eqnarray*}
\frac{\lambda}{\xi}&~=~&\frac{\lambda_0}{\xi_0}\Big(\frac{t}{t_0}\Big)^{\alpha}~=~\frac{3}{2}\Big[-\del_t\big(\frac{\dot{a}}{a}\big)+\frac{3\alpha\dot{a}}{2ta}+\frac{\alpha}{4t^2}(\alpha-2)\Big]\nonumber\\
\frac{\rho}{\xi}&~=~&\frac{3}{2}\Big[-\del_t\big(\frac{\dot{a}}{a}\big)+\frac{\alpha\dot{a}}{2ta}+\frac{\alpha}{4t^2}(\alpha+2)\Big]
\end{eqnarray*}
From the first equation above, the only power law solution for $\lambda\neq 0$ is given by: $\alpha=-2,~\beta=-4$. This results in the following expression for the matter density: $\frac{\rho}{\xi}=-\frac{3}{2t}\del_t\Big(\frac{t\dot{a}}{a}\Big)$, which identically vanishes for any power law. This simply reflects the vacuum solution, and hence should be ignored. In other words, the radiation epoch admits no solution with a nontrivial $\lambda$.

For $\lambda=0=\Lambda$, the field equations imply:
\begin{eqnarray*}
\frac{\rho}{\xi}~=~-3\Big[\frac{\ddot{a}}{a}+\frac{\alpha \dot{a}}{2ta}-\frac{\alpha}{2t^2}\Big]~=~3\Big[\Big(\frac{\dot{a}}{a}\Big)^2+\frac{\alpha \dot{a}}{ta}+\frac{\alpha^2}{4t^2}\Big]
\end{eqnarray*}
This has the following solution:
\begin{eqnarray}
a(t)~=~a_0\Big(\frac{t}{t_0}\Big)^{\frac{2-\alpha}{4}}
\end{eqnarray}
where $\alpha<2$ in an expanding universe. Note that $\rho=\frac{3}{4}\Big(1+\frac{\alpha}{2}\Big)^2\sim \frac{1}{t^{(2-\alpha)}}\sim \frac{1}{a^{4}}$, as expected from the local conservation law.

A few important properties of the radiation epoch are worth emphasizing:\\
a. As in the previous case, these solutions also fall into two classes: decaying $G(t)$ for $0<\alpha<2$ and growing $G(t)$ for $\alpha<0$.\\
b. Accelerating universe solutions correspond to a rapid variation with $\alpha<-2$. Thus, radiation domination admits only a decelerating universe for a relatively slow variation of the couplings. \\
c. For the age of a radiation dominated universe, we have $t_0\leq \frac{1}{2H_0}$ for a decaying $G(t)$ and $t_0> \frac{1}{2H_0}$ for a growing $G(t)$.\\
d. The vanishing of $\lambda$ (or, $\Lambda$) at the radiation epoch suggests that the universe could be characterized by either the same or different signs of $\lambda$ before and after this phase of evolution. This is not in conflict with an accelerating universe at present, since that only suggests that its sign is currently positive.

\section{Distance-Redshift relation}

Given the solutions obtained for any $w$, let us consider the radial coordinate distance traversed by light beginning from a source (let us say, from a standard candle such as SNe 1a) at a redshift $z$ till the current instant $t=t_0$:
\begin{eqnarray}
\chi(z)&~=~&\frac{1}{a_0}\int_0^z \frac{dz}{H(z)}~=~\frac{1}{H_0}\Bigg[\frac{(1+z)^{1-\frac{3(1+w)}{2-\alpha}}-1}{1-\frac{3(1+w)}{2-\alpha}}\Bigg]\nonumber\\
&~=~ & \frac{1}{a_0 H_0}\Big[z~-~\frac{3(1+w)}{2(2-\alpha)}z^2~+~\Bigg(\frac{1+w}{2(2-\alpha)}\Bigg)
\Bigg(\frac{3(1+w)}{(2-\alpha)}+1\Bigg)z^3~+~o(z^4)\Bigg]
\end{eqnarray}  
where in the last line we have expanded L.H.S. in a series around $z=0$. The first term at the R.H.S. reflects the Hubble's law. Clearly, for a given $w$, a precise measurement of the coefficients of the quadratic and cubic terms above could be used to determine the equation of state and the exponent of variation ($\alpha$).

Using the lower limit on the age of the universe: $t_0\gtrsim 1.3\times 10^{10}$ years, we find that the present epoch must correspond to $w\leq 0$ ($w<0$ if only a slow variation of couplings is allowed). Here we plot the comoving distance as a function of redshift in Fig.1 and Fig.2 and compare with the $\Lambda CDM$ prediction for the best fit parameters $\Omega_m\approx 0.32,~\Omega_\Lambda\approx 0.68$ (the thick black line) for the range $0\leq z\leq 1.5$. The dashed lines represent accelerated universe.
 The dust-dominated distances are seen to approach the $\Lambda CDM$ curve provided the variation of $\xi$ is fast enough ($\alpha< -\frac{3}{2}$). On the other hand, for $w=-\frac{1}{3}$, such a correspondence occurs for a much smaller variation: $\alpha\approx -\frac{1}{2}$. 
 Let us also observe that the dotted lines ($\alpha=1$), corresponding to a $G(t)$ according to Dirac's large number hypothesis, is not consistent with observations.

This suggests that for an accelerating universe now \cite{perl,riess}, both $w$ and $\alpha$ must be negative, provided the couplings do not vary too rapidly. Thus, a Newton's coupling that is growing in time at the present epoch is preferred. The slower the variation of the coupling, the farther away the effective EOS today is expected to be from $w=0$. 
 
\begin{figure}
\includegraphics[scale=.69]{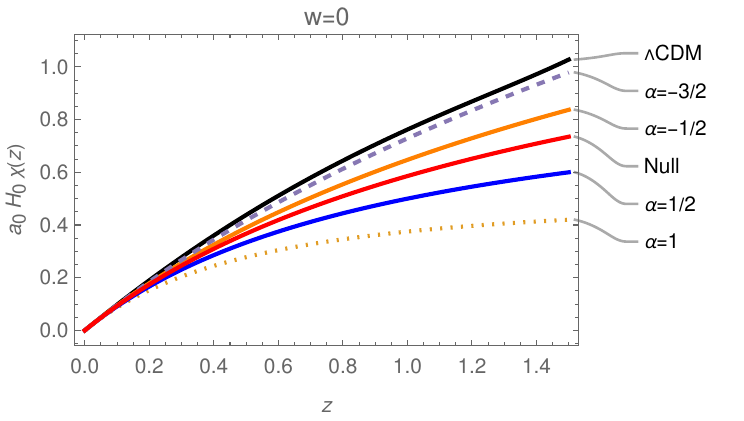}
\caption{Comoving distance for dust matter domination ($w=0$) for different positive and negative exponents $\alpha$ ($=1,\frac{1}{2},0,-\frac{1}{2},-\frac{3}{2}$). The null curve ($\alpha=0$) corresponds to a pure matter-domination ($\Omega_\Lambda=0$) in a $\Lambda CDM$ Universe. The thick black line refers to the $\Lambda CDM$ curve assuming $\Omega_m=0.32,\Omega_\Lambda=0.68$.}
\end{figure}
\begin{figure}
\includegraphics[scale=.67]{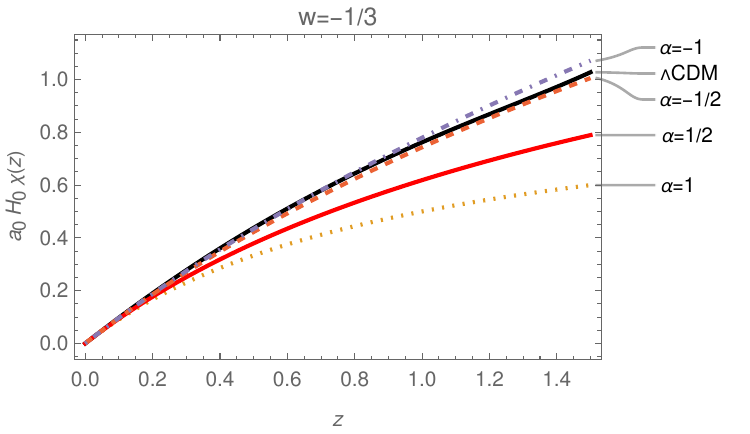}
\caption{Comoving distance for the EOS $w=-\frac{1}{3}$, plotted for different positive and negative exponents $\alpha$ ($=1,\frac{1}{2},-\frac{1}{2},-1$). $\alpha=1$ corresponds to Dirac's conjecture (LNH) $G\sim t^{-1}$, which is clearly disfavoured.}
\end{figure}

\section{Constraints on the variation of $G$ and $\Lambda$}
In this section, using the limits on the age of the Universe and assuming $\Lambda>0$, we present numerical bounds on the variation of both the couplings.

The rate of variation of the Newton's coupling at present reads: $\frac{\dot{G}}{G}=-\frac{\dot{\xi}}{\xi}=-\frac{\alpha}{t_0}$. Using the upper and lower bounds on the age, we find a dimensionless constraint as: $-\frac{\alpha H_0}{2}<\frac{\dot{G}}{G}<-\alpha H_0$. Assuming $w=-\frac{1}{3}$ at the present epoch and $\alpha\approx -\frac{1}{2}$ using the distance-redshift diagram, we find: 
\begin{eqnarray}
1.67\times 10^{-11} yr^{-1}<\frac{\dot{G}}{G}<3.34\times 10^{-11} yr^{-1},
\end{eqnarray}
where $H_0=67$ $km s^{-1}Mpc^{-1}$.
Note that this bound is stronger than the existing ones obtained within supernova cosmology assuming a $\Lambda CDM$ model and extending upto about $z\sim 1$ \cite{mould,gaztanaga}.

Using similar arguments, we obtain a new bound on the variation of the cosmological coupling as:
\begin{eqnarray}
-2H_0<\frac{\dot{\Lambda}}{\Lambda}<-H_0.
\end{eqnarray}
Unlike $G$, this constraint is independent of $\alpha$. This numerically translates to: $-6.7\times 10^{-11} yr^{-1}>\frac{\dot{\Lambda}}{\Lambda}>-13.4\times 10^{-11} yr^{-1}$.

\section{A solution to the Cosmic Coincidence Problem}
 
In the $\Lambda CDM$ model, one of the important unresolved issues is the apparent coincidence as to why the energy densities due to $\Lambda$ and the matter should roughly be of the same order of magnitude only at the present epoch. It should be emphasized that the ratio of these densities, which is typically inferred from the best fit data consistent with an accelerating universe now, is severely model-dependent.
Thus, for any model other than $\Lambda CDM$, a resolution of this problem would require the same temporal behaviour of the two densities during most of the period of evolution of the universe: $\rho_m\sim \rho_\Lambda$, so that their ratio is a constant (which could be different in general from the best fit 
$\Lambda CDM$ data) for a given epoch \cite{manohar} and do not vary over many orders of magnitudes between epochs.

Let us now observe that unlike standard FLRW cosmology, the effective $\Lambda$ here is directly related to the energy momentum tensor through the equation of motion (\ref{c-eom2}). Further, the solutions obtained show that the ratio of the energy densities related to (baryonic and non-baryonic) matter and $\Lambda$ is a constant for a given epoch. As we shall now show, nor is the current epoch a special one (assuming it being dominated by an effective equation of state $w\leq 0$), neither is there a coincidence regarding the relative order of magnitudes of these densities within this theory.

First, keeping in mind the standard parametrization, let us rewrite the equation of motion (\ref{c-eom}) as:
\begin{eqnarray}
\frac{\rho}{3\xi H_\xi^2}+\frac{\lambda}{3\xi H_\xi^2}=1
\end{eqnarray}
where $H_\xi\equiv\frac{\dot{a}}{a}+\frac{\dot{\xi}}{2\xi}$ naturally emerges as a new Hubble variable superceding $H\equiv \frac{\dot{a}}{a}$. The fractional density components corresponding to matter and $\lambda$ are then given by:
\begin{eqnarray}
\Omega_m=\frac{\rho}{3\xi H_\xi^2},~\Omega_\lambda=\frac{\lambda}{3\xi H_\xi^2}\equiv 1-\Omega_m
\end{eqnarray}
Using the generic solutions found earlier, we obtain:
\begin{eqnarray}
\Omega_\lambda=-\frac{\alpha(1-3w)}{4+\alpha(1+3w)}=-\frac{\alpha(1-3w)}{2(\alpha+2)}\Omega_m
\end{eqnarray}

Given the broad range of the current limits on $\frac{\dot{G}}{G}$ presented within various contexts \cite{ujan,*uzan,ajith}, it might be premature to decide whether the exponent of variation $|\alpha|$ lies closer to $0$ or is of order unity. Hence, let us consider both possibilities as below.

As confirmed by observations, let us assume that the universe has entered a phase of accelerated expansion after the epoch of matter domination ($\lambda>0,~w\lesssim 0$). Given the inequality $\Omega_m>\Omega_\lambda$, we have: $\alpha>-\frac{4}{3}$ for a lower bound of the exponent during matter domination. Hence, we may assume that $\alpha$ is close to $-1$ ($>-1$ for decelerating universe) during this epoch, for which the ratio becomes: $\frac{\Omega_m}{\Omega_\lambda}|_{w=0}\sim 2$.

Consistency of the $a_0 H_0 \chi(z)$ vs $z$ diagram (figure $2$) with the supernovae data, however, indicates that for a slower variation of $\alpha$ than above ($-1<\alpha<0$), an accelerated universe now is consistent with a negative EOS $0>w\geq -\frac{1}{3}$. The corresponding ratio then becomes: $\frac{\Omega_m}{\Omega_\lambda}|_{w=-\frac{1}{3}}=|-\frac{\alpha+2}{\alpha}|=\frac{1}{2}\frac{\Omega_m}{\Omega_\lambda}|_{w=0}$. Thus, for a given $\alpha$, the density ratio remains of a similar order of magnitude between the current and matter dominated epoch, which define most of the lifetime of the universe. 

Finally, let us also consider the question if the ratio could be widely different in order of magnitude during the early stages of evolution. Since $w>\frac{1}{3}$ is expected to dominate the density at early times in an expanding universe, we may take the stiff EOS as an extreme representative of the early phase. A positive $\lambda$ and a relatively slowly varying $G(t)$ implies $0<\alpha<2$. In this epoch we obtain: ${\frac{\Omega_m}{\Omega_\lambda}}=\frac{\alpha+2}{\alpha}$, which, again, is of a similar order of magnitude as in the current and matter dominated epochs irrespective of whether $|\alpha|$ is closer to 0 or 1.
 
Thus, within the framework here, the ratio of densities is predicted to be of similar order of magnitudes for about the whole range of equation of state (except the critical epoch of radiation which, owing to the associated density fall-off occupies only a small duration of the total period of evolution). Consequently, the current epoch cannot be considered to be a special one in terms of different density components.

Thus, the action formulation (\ref{S}) rids our universe of the so-called coincidence problem involving $\Omega_m$ and $\Omega_\lambda$ ($\Omega_\Lambda$).



\section{Concluding remarks}

We have set up a Lagrangian formulation of gravity with matter where all the gravitational couplings could be treated universally, as being dependent on a single scalar field. We find the most general power law solutions for FLRW spacetimes and the possible class of time-evolution of the couplings. Our analysis provides a basis for setting up phenomenological models in the context of variable gravitational couplings. 

There are a few significant imports of our results here. The effective $\Lambda$, superceding the `dark energy' component, becomes directly related to the energy momentum tensor of the matter fluid through the field equations. The radiation dominated epoch emerges as a critical phase where $\Lambda$ vanishes. During any other epoch ($w\neq\frac{1}{3}$), the matter and $\Lambda$ densities exhibit a constant ratio, in contrast with the standard $\Lambda$CDM model based on constant couplings. This reflects a new method to solve the cosmic coincidence problem involving these two densities dynamically, and could supercede the quintessance models or their variants \cite{peebles,zlatev}.
In principle, our formulation admits the possibility of an accelerated expansion both in the early and late universe. 
These features show that the scope of the analysis here goes beyond  FLRW or the standard scalar-field based cosmology, and could be used to address some of the outstanding problems in these contexts.


It is intriguing to observe that $\xi\sim t$ emerges as a particular solution of the general class, resembling the time-evolution of $G$ as conjectured in Dirac's large number hypothesis (LNH). However, the critical differences between our formulation and the latter are obvious. To emphasize, the action principle here describes the time evolution of both $G$ and $\Lambda$ while being based on a single metric (the generalization to include any other gravitational coupling is straightforward), and leads to no exotic feature such as a spontaneous creation of matter in empty space, in contrast with the LNH. 

While the resulting distance-redshift relations are shown to rule out the LNH for the present universe, a growing $G(t)$ ($\alpha<0$) along with a negative EOS is preferred at late times. This prediction for the Newton's coupling is new and intriguing. 
Further, we have improved upon the bound on $\dot{G}$ in the context of supernova cosmology, and have obtained a new bound on $\dot{\Lambda}$. 

\acknowledgments 
The support of the MATRICS grant MTR/2021/000008, SERB (ANRF), Govt. of India is gratefully acknowledged.

 \bibliography{coincidence-jcap}

\end{document}